\documentclass[12pt,preprint]{aastex}
\usepackage{epsfig}
\begin{document}

\small

\title{Candidate Type II Quasars from the SDSS: III. Spectropolarimetry Reveals Hidden Type I Nuclei\footnote{The observations reported here were obtained at the Multiple Mirror Telescope Observatory, a facility operated jointly by the Smithsonian Institution and the University of Arizona, and with the NASA/ESA Hubble Space Telescope.}}

\author{
Nadia L. Zakamska\altaffilmark{2}, 
Gary D. Schmidt\altaffilmark{3},
Paul S. Smith\altaffilmark{3}, 
Michael A. Strauss\altaffilmark{2}, \\
Julian H. Krolik\altaffilmark{4},
Patrick B. Hall\altaffilmark{2,5}, 
Gordon T. Richards\altaffilmark{2}, 
Donald P. Schneider\altaffilmark{6},\\
J. Brinkmann\altaffilmark{7},
Gyula P. Szokoly\altaffilmark{8}
\altaffiltext{2}{Princeton University Observatory, Princeton, New Jersey 08544}
\altaffiltext{3}{Steward Observatory, The University of Arizona, 933 North Cherry Avenue, Tucson, AZ 85721}
\altaffiltext{4}{Department of Physics and Astronomy, Johns Hopkins University, 3400 North Charles Street, Baltimore, MD 21218-2686}
\altaffiltext{5}{Department of Physics and Astronomy, York University, 4700 Keele Street, Toronto, Ontario, M3J 1P3, Canada}
\altaffiltext{6}{Department of Astronomy and Astrophysics, Pennsylvania State University, University Part, PA 16802}
\altaffiltext{7}{Apache Point Observatory, P.O. Box 59, Sunspot, NM 88349} 
\altaffiltext{8}{Max-Planck-Institut f\"ur extraterrestrische Physik, Giessenbachstra{\ss}e 1, D-85741 Garching, Germany} 
}

\begin{abstract}
We have conducted spectropolarimetry of 12 type II (obscured) quasar candidates selected from the spectroscopic database of the Sloan Digital Sky Survey based on their emission line properties. Polarization was detected in all objects, with nine being highly polarized ($>3\%$) and with polarization reaching as high as 17\% in two objects. Broad lines were detected in the polarized spectra of five objects. These observations prove beyond a reasonable doubt that the objects in our sample are indeed type II quasars, in that they harbor luminous UV-excess AGNs in their centers and that the direct view to the AGN is highly obscured. For three of the objects in this paper, we have obtained HST images in three bands. The HST observations, combined with the spectropolarimetry data, imply that scattering off material outside the obscuration plane is the dominant polarization mechanism. In all three objects the sizes of scattering regions are a few kpc. For one object, the extent of the scattering region, coupled with the characteristics of the polarized spectrum, argue strongly that dust scattering rather than electron scattering dominates the polarized light. Our observations are well-described by the basic orientation-based unification model of toroidal obscuration and off-plane scattering, implying that the model can be extended to include at least some high-luminosity AGNs.
\end{abstract}

\keywords{galaxies: active --- polarization --- quasars: general --- quasars: emission lines}

\section{Introduction}
Unification models of active galactic nuclei (AGNs) aim to explain the differences in observed AGN properties as orientation effects due to the non-isotropic nature of obscuring material \citep{anto93}. In these models, the observer's view to the central engine can be blocked along some lines of sight by very optically thick, toroidally concentrated dusty material, making UV, optical and soft X-ray emission from AGNs highly anisotropic. A strong UV continuum and broad permitted emission lines that originate very close to the central engine are typical of sight-lines without significant obscuration (type I AGNs). These features are not observed in unpolarized optical/UV light in sources with large amounts of obscuration along the line of sight (type II AGNs).

Unification models are well established for low-luminosity, nearby AGNs (Seyfert galaxies); whether they can be extended to the luminosity and redshift range of genuine quasars has long been controversial. Type II quasars are the luminous analogs of Seyfert 2 galaxies, but despite their high intrinsic luminosities they have been very hard to find, and only a handful of candidates have been described in the literature \citep{klei88,mcca93,hine95,daws01,djor01,norm02,schm02,ster02,dell03,smit03}.  

We have identified about 150 type II quasar candidates in the redshift range $0.3<z<0.8$ from the spectroscopic data of the Sloan Digital Sky Survey (\citealt{z03}, hereafter Paper I). Selecting such a large number of these optically faint objects is possible because of the size of the spectroscopic survey (currently containing $3.7 \times 10^5$ publicly available spectra; \citealt{abaz04}), part of which is dedicated to the search for unusual objects. We are now conducting a multi-wavelength follow-up campaign of the most luminous objects in our sample to constrain physical models of the circumnuclear obscuration and its geometry, and to probe the unification model at high luminosities.   

The first strong evidence for the unification model of AGNs came from spectropolarimetry of Seyfert galaxies pioneered by \citet{anto85} for NGC1068. In many Seyfert 2 galaxies, while the direct emission from the AGN is obscured, some of the light from the central engine and the broad-line region is scattered by material outside the plane of the obscuring torus and reaches the observer. Because scattered radiation is polarized, observations of the polarized flux can thus offer an indirect view of the central parts of the obscured AGNs. 

In this paper we present spectropolarimetry of twelve type II quasar candidates, probing the central parts of these AGNs otherwise hidden from the observer.  In Section \ref{sec_data} we describe the observations and data reduction. In Section \ref{sec_results} we summarize the results of the spectropolarimetry observations and present example spectra. We discuss different mechanisms of producing polarization in AGNs in Section \ref{sec_discussion}, along with related observations obtained by the Hubble Space Telescope (HST). We summarize our results in Section \ref{sec_conc}. An $h=0.7$, $\Omega_m=0.3$, $\Omega_{\Lambda}=0.7$ cosmology is assumed throughout. Optical emission and absorption lines are identified by their air wavelengths in \AA\ (e.g., [OIII]5007). Objects are identified by their J2000 coordinates in Table 1 (e.g., SDSS J084234.94+362503.1) and shortened to $hhmm+ddmm$ notation  in the text (SDSS~J0842+3625).

\section{Sample selection, observations and data reduction}
\label{sec_data}

The Sloan Digital Sky Survey (SDSS; \citealt{york00, stou02, abaz03, abaz04}) uses a drift-scanning imaging camera \citep{gunn98} to image 10,000 deg$^2$ of the sky in the SDSS $ugriz$ AB magnitude system \citep{fuku96, lupt99}. The astrometric calibration is better than $0\farcs 1$ rms per coordinate \citep{pier03}, and the photometric calibration is accurate to 3\% or better \citep{hogg01, smit02j}. Using a 640 fiber double spectrograph, the SDSS carries out spectroscopy ($3800-9200$\AA\ with a resolution of $\sim 2000$) of complete samples of galaxies \citep{stra02}, luminous red galaxies \citep{eise01} and quasars \citep{rich02}. A small number of other types of sources are targeted \citep{stou02} depending on fiber availability \citep{blan03}.

As described in Paper I, we searched the SDSS spectroscopic database as of September 2002 for type II AGNs, which are characterized by narrow ($FWHM<1000$ km sec$^{-1}$) emission lines. The absence of broad permitted lines, very high [OIII]5007/H$\beta$ ratios and the presence of high-ionization emission lines ([NeV]3346,3426) distinguish type II AGNs from star-forming galaxies and broad-line AGNs. The source of the ionizing radiation and the broad-line region are obscured and are not seen in the optical spectra. Based on these characteristics, we identified 291 objects in the redshift range $0.3<z<0.8$. We used the [OIII]5007 emission line as a proxy for the intrinsic luminosity, as it is emitted from the extended, and therefore presumably largely unobscured, narrow-line region; we showed that luminous quasars ($M_B<-23$) typically have [OIII]5007 line luminosities in excess of $3\times 10^8 L_{\odot}$. Hence we use this value to distinguish between type II quasars and Seyfert 2 galaxies, and by this criterion about 50\% of the objects in our sample are luminous enough to be classified as type II quasars. The infrared luminosities of the objects in the sample, as detected by IRAS, reach $3\times 10^{46}$ erg sec$^{-1}$, placing them among the most luminous quasars at the redshifts of our sample (\citealt{z04}, hereafter Paper II). 

From the sample of 291 type II AGNs, we selected several of the most luminous objects in the [OIII]5007 emission line ($L$([OIII]5007)$\ga 10^9 L_{\odot}$) for spectropolarimetric observations. Spectropolarimetry was obtained between October 2003 and May 2004 for twelve objects (Table 1). We used the CCD Spectropolarimeter (SPOL, \citealt{schm92b}) at the 6.5m Multiple Mirror Telescope (MMT). All observations used a low-resolution grating providing spectral coverage of 4100-8200\AA. An entrance slit of 1.1\arcsec\ width was used, resulting in a spectral resolution of $\sim$19\AA. The slit was aligned with atmospheric dispersion and rotator set to track field rotation.

Calibration procedures included observing with a completely polarizing prism in the beam to measure the instrumental efficiency. The measured polarization position angles (given in degrees east of north) were referenced to the equatorial system by observing interstellar polarization standard stars \citep{schm92a}, and the low instrumental polarization ($\la 0.1\%$) of SPOL was confirmed by observing unpolarized stars \citep{schm92a}. The degree of polarization is quantified by the rotated Stokes parameter $q(\lambda)=[Q(\lambda) \cos 2\theta + U(\lambda) \sin 2\theta]/I(\lambda)$, where $\theta$ is the mean polarization angle of the source in the 5000-8000\AA\ bandpass and $I(\lambda)$ is the total intensity. For wavelength-independent $\theta$ the rotated Stokes parameter is mathematically indistinguishable from the conventional polarization $P=\sqrt{Q^2+U^2}/I$ (since the polarization angle is defined as $\tan 2\theta=U/Q$), but unlike the positively defined $P$, the parameter $q$ has a virtually normal error distribution. All our candidates are at high Galactic latitudes ($|b|>30^{\circ}$), so the Galactic interstellar polarization is expected to be less than 0.8\% in our objects (\citealt{berr90, impe90} and Section \ref{sec_broad_band} below). 

As a byproduct of our spectropolarimetry program, our MMT SPOL observations provided us with optical spectra which have much higher signal-to-noise ratios (S/N) than the original SDSS spectra because of the collecting power of the 6.5 m telescope, although at somewhat lower spectral resolution. The S/N gain is very noticeable in the continuum, which in our sample of type II AGNs is typically a few $\times 10^{-17}$ erg sec$^{-1}$ cm$^{-2}$ \AA$^{-1}$, comparable to the spectral flux errors per pixel in the SDSS spectra. The total fluxes measured by MMT are affected by seeing and the sizes of the sources relative to the 1.1\arcsec-wide slit, and are typically less than those from the SDSS 3\arcsec\ fiber spectra. Flux calibration of the SPOL data is achieved using observations of flux standards obtained with the same instrumental setup. The MMT observations are typically referenced to observations of two standard stars each night selected from the list of \citet{mass88}. Additional observations of A and B-type SAO stars immediately following and near each AGN target enabled removal of terrestrial absorption features. Three AGNs were observed on more than one night, from which a measure of the consistency of the flux calibration could be assessed. From these results, we estimate that the relative (blue/red) spectral energy distributions are accurate to $\pm$5\% over the range $4500-7500$~\AA, degrading somewhat at the wavelength extremes. Absolute flux levels for the spectra are reliable to $\pm15$\%. For clarity, we refer to the MMT or the SDSS optical spectra as `total flux' spectra ($F_{\lambda}$), in contrast to the `polarized flux' MMT spectra ($q\times F_{\lambda}$).

Three of the objects discussed in this paper (Table 1) have been imaged by HST as part of our program ``The Host Galaxies of Type II Quasars'' (program ID 9905). HST observations consisted of imaging with the Wide Field Channel of the Advanced Camera for Surveys (ACS) in three bands centered near rest-frame 3200\AA, 4000\AA\ and 5600\AA\ (for SDSS~J1323$-$0159) or 6600\AA\ (for SDSS~J1039$+$6430 and SDSS~J1413$-$0142). These bands were chosen to avoid strong emission lines such as [OII]3727 and [OIII]4959, 5007 and therefore minimize the contribution from the AGN emission to the images. All data reduction was done using the Multidrizzle routine available as part of the Space Telescope Science Data Analysis System v.3.2 \citep{koek02}. The images in three bands were then combined to produce RGB composite images using the routine by \citet{lupt04}. Further details on the HST observations, data reduction and data analysis will be presented in our subsequent paper (\citealt{z05}, hereafter Paper IV). In Section \ref{sec_discussion} we discuss the HST observations in relation with the spectropolarimetry program. 

\section{Optical polarization}
\label{sec_results}

\begin{figure}
\epsscale{0.8}
\plotone{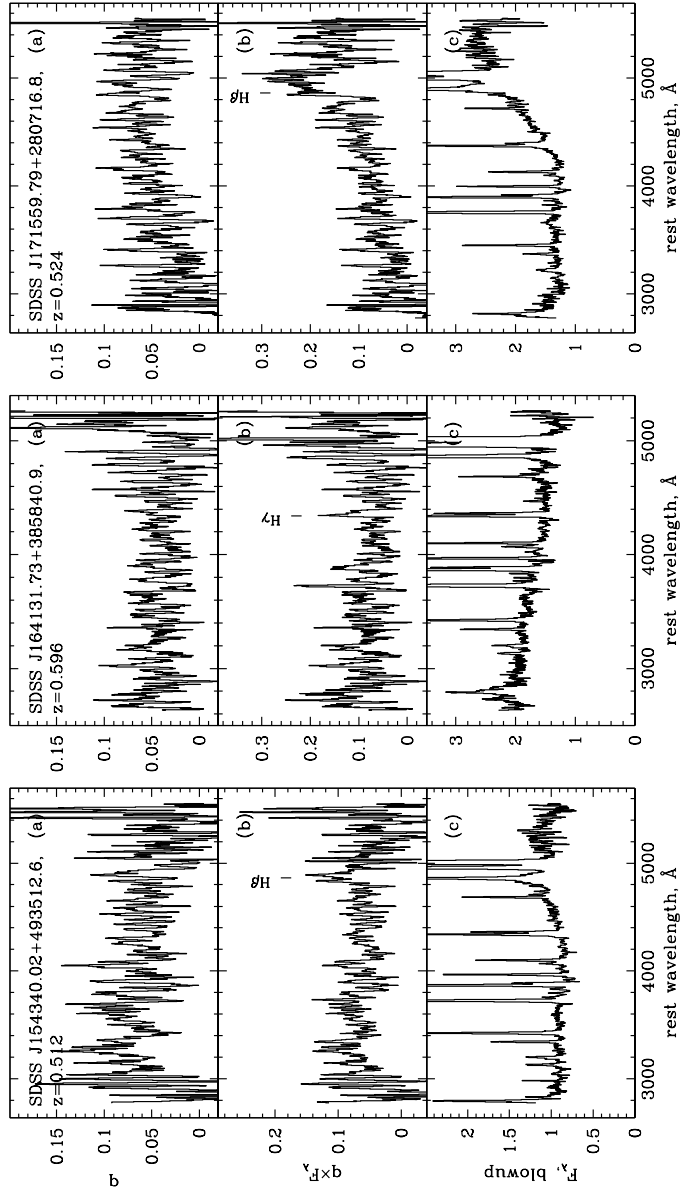}
\figcaption{\small SPOL spectra of SDSS~J1543+4935 (left), SDSS~J1641+3858 (middle), and SDSS~J1715+2807 (also known as 2MASS J171559.7$+$280717, right). From top to bottom: (a) polarization fraction smoothed by 5 pixels (7 pixels for SDSS~J1641+3858), (b) polarized flux density smoothed by 5 pixels (7 pixels for SDSS~J1641+3858), (c) total optical flux density. Flux densities in panels (b) and (c) are in units of $10^{-17}$ erg sec$^{-1}$ cm$^{-2}$ \AA$^{-1}$. Broad H$\beta$ is detected in the polarized spectrum in SDSS~J1543+4935 and in SDSS~J1715+2807, and broad H$\gamma$ is detected in SDSS~J1641+3858. \label{pic_broad}}
\end{figure}

Our spectropolarimetry program on the MMT was performed in two stages. Each object was initially observed in a relatively short (0.5$-$1 hour) exposure to determine the overall level of polarization. Then we re-observed objects with high polarizations and coadded all available exposures to search for spectral features in the polarized spectra. Since our objects are typically faint in the continuum ($F_{\lambda}=10^{-17}-10^{-16}$ erg sec$^{-1}$ cm$^{-2}$ \AA$^{-1}$, or about $m_{AB}=21$ mag in the red), emission lines in the polarized spectrum can be detected in a reasonable time only for sources with high polarization. The main results of our program are summarized in Table 1, and example spectra are presented in Figures \ref{pic1}-\ref{pic_subtraction}. 

\subsection{Broad-band polarization}
\label{sec_broad_band}

In Table 1, the average polarization calculated over the entire spectral range of SPOL, $\langle q \rangle=\int q d\lambda/(\lambda_{max}-\lambda_{min})$, is given by `mean $q$'. `Blue $q$' is the mean value of the continuum polarization in the rest-frame range 2820$-$3710\AA, between the MgII 2800 and the [OII]3727 emission lines (or just blueward of the [OII]3727 emission line if only part of the 2820\AA$-$3710\AA\ rest-frame range is observed due to the SPOL wavelength coverage).  Similarly, `green $q$' is the mean polarization in the rest-frame range 4000$-$4840\AA\ (between [OII]3727 and H$\beta$) and `yellow $q$' is the mean polarization redward of the [OIII]5007 emission line (5030$-$6530\AA). Narrow emission lines in our objects typically have low polarization, so we chose these three bands to avoid the strongest lines (Figure \ref{pic1}). We also masked out all weaker narrow emission lines when calculating the mean polarization values in these bands. 

From Table 1, the immediate result is that the average polarization values are very high: out of 12 objects, 9 show mean polarization above $P=3\%$, the traditional high-polarization criterion for quasars \citep{moor84}. For comparison, none of the Palomar-Green quasars \citep{schm83} show polarization in excess of 2.5\%, and the mean polarization level of the entire sample is just 0.5\% \citep{berr90}. A large difference in the polarization levels between type I and type II quasars is expected in AGN unification schemes, since in unobscured quasars the strong UV and optical continuum dilutes the polarized component, producing small fractional polarization. The high polarization values strongly support our interpretation of the SDSS type II quasar candidates as obscured sources.

The polarization of SDSS type II quasar candidates is also much stronger than in red 2MASS AGNs \citep{cutr02}, of which only only 10\% exhibit $P>3\%$ \citep{smit02p}. This fact suggests that in many red 2MASS AGNs the direct light from the nucleus, although heavily reddened, makes a significant contribution to the optical continuum and dilutes the strongly polarized component, whereas the direct component is negligible in SDSS type II quasars. In fact, the only object that shows any evidence of a direct AGN component of our sample of 12 objects is SDSS~J1715+2807 (Figure \ref{pic_broad}, right). This object is one of the red 2MASS AGNs found by \citet{cutr02}, but it was also recovered independently from the SDSS data as a type II quasar candidate (Paper I). Its optical spectrum shows a weak but prominent red continuum which cannot be attributed to the host galaxy, as it does not show stellar absorption lines or the 4000\AA\ break. From HST $I$-band photometry of this object, \citet{marb03} estimate that only up to half of the red light can be attributed to the stellar component. No featureless red continuum has been convincingly detected in the total light of the other objects presented in this paper, and we conclude that the direct reddened light from the AGNs is not a significant contribution to the continuum spectra of most SDSS type II quasars. 

The unpolarized stellar light from the host galaxies of AGNs can significantly dilute the polarized component, reducing the measured fractional polarization. The `blue' and `green' range polarizations may be diluted by light from a young stellar population, whereas the `green' and the `yellow' ranges may be affected by old stars. It is common to subtract a template galaxy spectrum in order to determine the polarization fraction of the AGN itself. If $h(\lambda)$ is the fraction of the light contributed by the host galaxy at wavelength $\lambda$, then the polarization fraction at this wavelength corrected for the starlight dilution is  $q^*(\lambda)=q(\lambda)/(1-h(\lambda))$. However, the properties of the stellar population that contributes to the total optical spectrum are usually quite uncertain, and different host galaxy templates can lead to drastically different starlight corrections \citep{tran95a}. 

With these uncertainties in mind, we have attempted to correct the observed total MMT spectra (and therefore the polarization fraction) for starlight. The MMT total optical spectra are of sufficient S/N that the continuum is well detected in all cases. The inferred host fraction is very sensitive to the host galaxy template (e.g., old stellar population, young stellar population, or starburst), and even for a given template the range of the stellar contribution allowed by the observations is large. We restricted ourselves to two templates: (i) an optical and UV elliptical galaxy template by \citet{kinn96} to represent the old stellar population; and (ii) an A0V-star optical and UV template by \citet{pick98} to represent a young stellar population. Both templates were smoothed to have the resolution of our MMT spectra, which is low enough that the stellar velocity dispersion of the host galaxy is immaterial. We then subtracted these templates (or some combination thereof) from the MMT spectra. Host galaxy subtraction was considered successful if the strong absorption features (CaII 3933, 3968, G-band 4300, Mg 5175) and the 4000\AA\ break (or 3700\AA\ break representative of the young stellar population) were reduced to the noise level. 

\begin{figure}
\epsscale{0.8}
\plotone{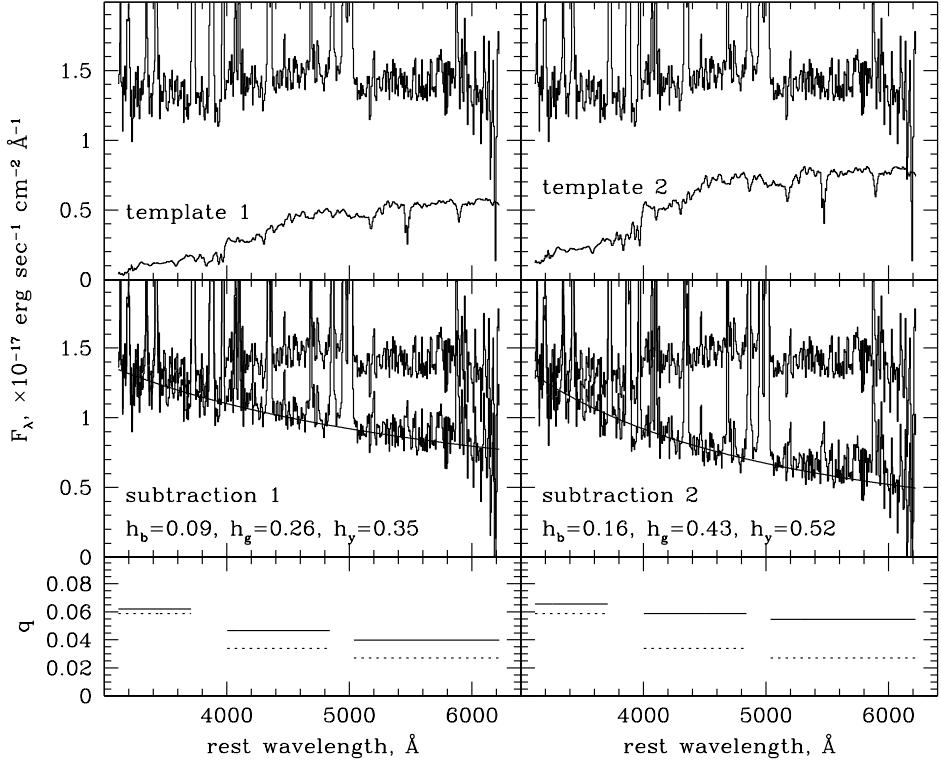}
\figcaption{\small Subtraction of the host galaxy contribution for SDSS J132323.33$-$015941.9 at $z=0.350$. The top panels show the total optical spectrum (top spectrum) and the template used for host subtraction (bottom spectrum). Templates 1 and 2 (T1 and T2, left and right) differ in the total amount of stellar light contribution and the composition: T1 is a pure elliptical galaxy spectrum and T2$=$T1$\times 1.3+$A0V. The middle panels show the total optical spectrum (top spectrum) and the result of the subtraction (bottom spectrum). The 4000\AA\ break disappeared as the result of subtraction for both templates, but the slope of the resulting continuum is different ($\alpha_{\nu}=-1.2$ for T1 and $\alpha_{\nu}=-0.6$ for T2, power-law fits shown as smooth lines). The bottom panels show the measured polarization fraction in the three bands in dotted lines and the dilution-corrected polarization fraction in solid lines. When corrected using T2, the polarization fraction is consistent with being wavelength-independent and the result of the subtraction has power index similar to that of the quasar continuum ($\alpha_{\nu}=-0.44$, \citealt{vand01}). The median galaxy fractions in three bands (blue, green and yellow) are listed in the middle panels for each case.  \label{pic_subtraction}}
\end{figure}

In Figure \ref{pic_subtraction} we illustrate typical uncertainties in the estimate of the host galaxy fraction by showing template subtraction for SDSS~J1323$-$0159. In Figure \ref{pic_subtraction}, left, we show the result of subtraction of an old stellar population contribution alone (represented by an elliptical galaxy template), and in Figure \ref{pic_subtraction}, right, we subtract a combination of an elliptical galaxy template and an A-star template. In this particular object the stellar absorption lines are not very obvious, but the 4000\AA\ break is clearly present. As this example shows, the host galaxy fraction is large, but can typically be determined only within a factor of 1.5, which introduces a large uncertainty in the dilution-corrected polarization level. In Table 1 we list our best estimates of the maximum amount of stellar light contribution allowed by the spectroscopic data; because of the large uncertainties, we give only one significant digit. We did not require that the result of the subtraction resemble a power-law, and in many cases it cannot be fit by a power-law as well as in SDSS~J1323$-$0159. 

The largest uncertainty in the host fraction is in the blue part of the spectrum, as it is impossible to calculate the contribution from young stars to the spectrum from the spectroscopic data alone. Balmer absorption lines and the Balmer break characteristic of young stars and the emission from the associated HII regions can be completely swamped by the high-luminosity emission from the narrow-line region of the AGN \citep{heck97}. Because of these difficulties, we did not add any more stellar or galaxy templates to our fitting procedure. The uncertainty in the spectrophotometric calibration of the spectra may introduce an additional uncertainty in the estimated host galaxy fraction, but it is typically smaller than that resulting from the fitting procedure. In Table 1 we list host fractions only for the `yellow' band, which is not severely affected by the contribution from young stars. Luminosities of subtracted templates are less than or comparable to $L^*$ in all cases except SDSS~J1501+5455; we further comment on the latter object in Section \ref{sec_ind_1501}.

The Galactic interstellar polarization has been shown to be $\la 9\%\times E(B-V)$ \citep{serk75}. For our objects Galactic reddening varies from 0.01 mag to 0.09 mag \citep{schl98}, so the interstellar polarization in the Galaxy is less than 0.8\% for all objects. Given very high values of polarization for most of our sources, the interstellar polarization is of concern only for the least polarized sources in the sample, SDSS~J1407+0217 and SDSS~J1501+5455. The reddening values toward these sources are 0.03 and 0.01, so even in these two sources the interstellar polarization ($\la 0.3\%$ and $\la 0.1\%$) cannot account for the observed polarization level. These two sources have the largest estimated host galaxy fraction of all the objects in our sample, so it is likely that their dilution-corrected polarization is much higher than observed.

\subsection{Polarized continuum}
\label{sec_pol_cont}

We have measured power-law indices of the polarized continuum ($F_{\nu}\propto\nu^{\alpha_{\nu}}$) for the three objects with the highest S/N of the polarized component: $\alpha_{\nu}=-0.3$ for SDSS~J0815+4304, and $\alpha_{\nu}=0.0$ for SDSS~J0842+3625 and SDSS~J1039+6430. The uncertainties in these measurements are about 0.2. Presumed unobscured quasars show slopes with power indices in the range $\alpha_{\nu}=-0.44\pm 0.3$ (\citealt{rich80, fran96, vand01, rich03}, although increasing amounts of intrinsic reddening make the distribution of indices non-Gaussian on the red end; \citealt{hopk04}), and the power indices of these three objects are consistent with unobscured quasar continuum slopes. In the objects SDSS~J1543+4935 and SDSS~J1715+2807 (Figure \ref{pic_broad}) the polarized spectrum is significantly redder than a typical unobscured quasar continuum, with $\alpha_{\nu}\simeq-2.0$ and $\alpha_{\nu}\simeq-4.0$, respectively (each with an uncertainty of about 0.5). In the remaining objects the S/N is too low to fit a power law. 

\subsection{Emission lines}

\begin{figure}
\epsscale{0.75}
\plotone{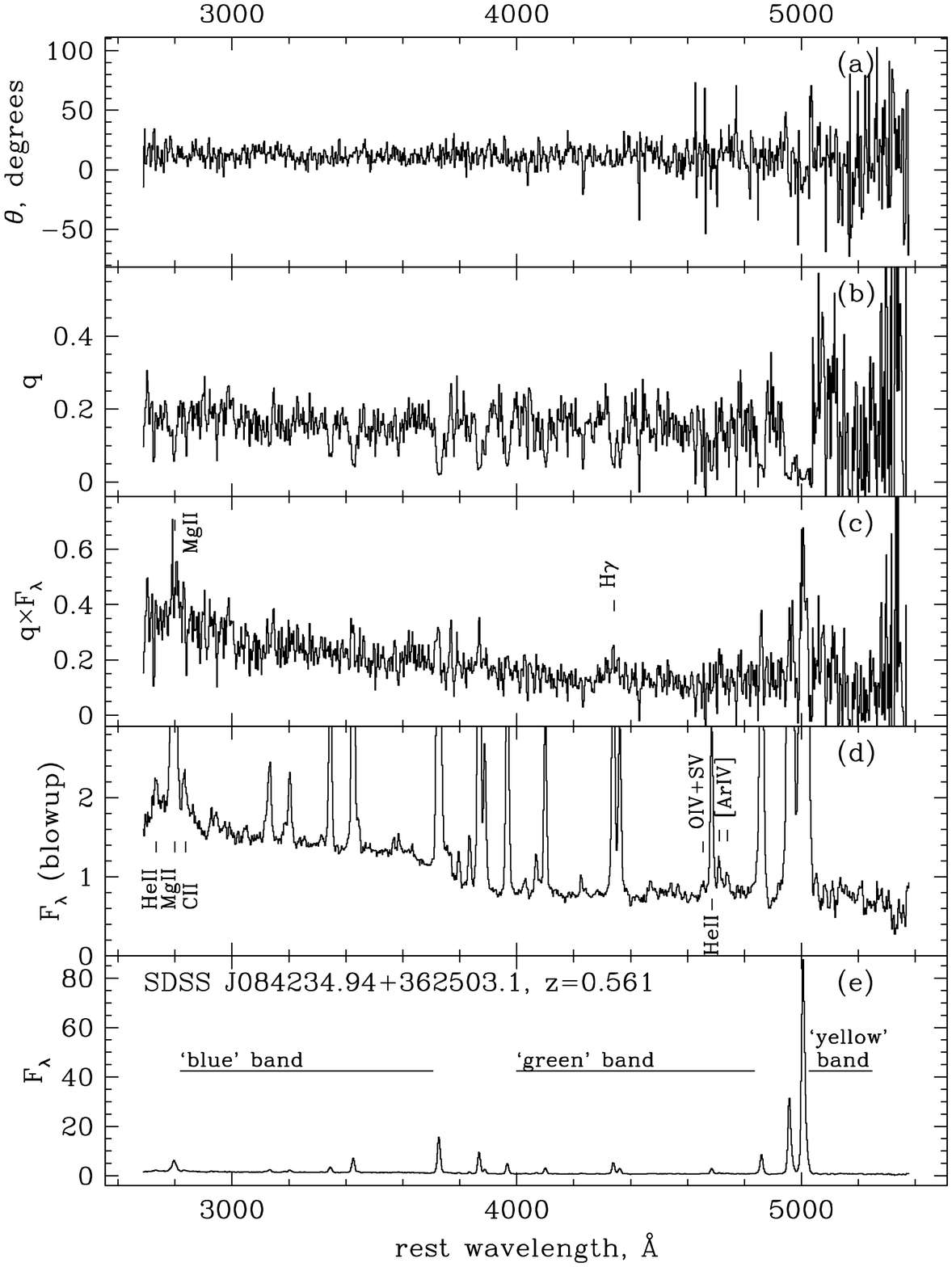}
\figcaption{\small SPOL spectrum of SDSS~J0842+3625: (a) polarization position angle in degrees; (b) polarization fraction (rotated Stokes parameter); (c) polarized flux density; (d) blowup of the optical spectrum to see the continuum shape; (e) total flux density. All flux densities (c,d,e) are in units of $10^{-17}$ erg sec$^{-1}$ cm$^{-2}$ \AA$^{-1}$; the spectral resolution is about 19\AA. All spectra presented in this paper (the total optical spectra and the polarized spectra) have been corrected for Galactic reddening using the analytical approximation to the extinction law by \citet{odon94} and the reddening values from \citet{schl98}. This object has the highest [OIII]5007 luminosity in the sample presented in Paper I; it shows wavelength-independent continuum polarization of about 16\%, among the highest ever observed in type II AGNs. Broad H$\gamma$ is detected in the polarized spectrum, as well as several narrow lines (MgII 2800, H$\beta$ and [OIII]4959, 5007). The polarized flux density is well-fit by a power-law with $\alpha_{\nu}=0.0\pm0.2$. \label{pic1}}
\end{figure}

\begin{figure}
\epsscale{0.75}
\plotone{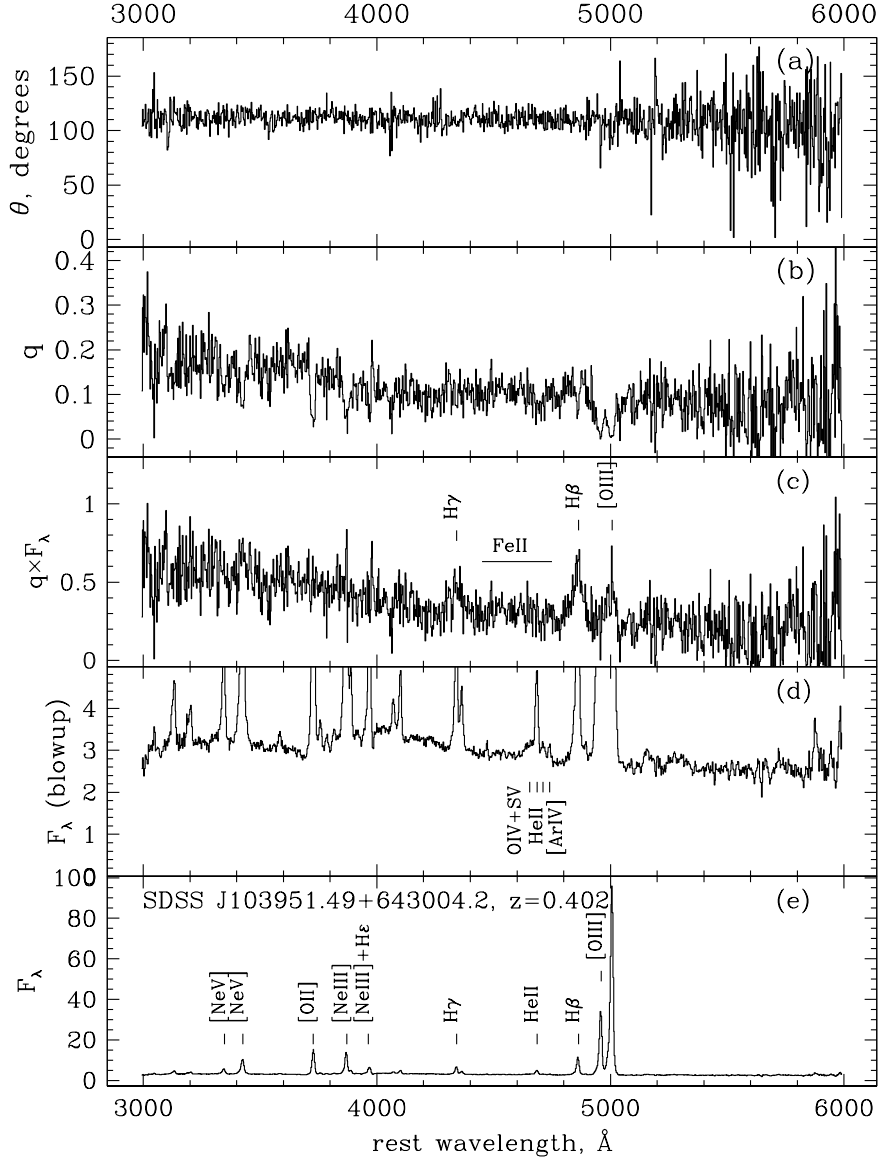}
\figcaption{\small SPOL spectrum of SDSS~J1039+6430; the panels and units are the same as in Figure \ref{pic1}. This object is one of the three objects from our spectropolarimetry sample that have been imaged in three bands by HST (Table 1, Paper IV, Figure \ref{pic_hst}). Broad Balmer lines (H$\beta$, H$\gamma$) are present in the polarized spectrum of this object, as are narrow [OIII]4959, 5007. In addition, there is a possible detection of FeII emission between H$\gamma$ and H$\beta$. The polarized flux density is well-fit by a power-law with $\alpha_{\nu}=0.0\pm0.2$. \label{pic2}}
\end{figure}

Broad emission lines in the polarized light were detected in the five objects shown in Figures \ref{pic1}-\ref{pic_broad}. The full widths at half maximum for these lines listed in Table 1 are based on Gaussian fits (the line widths are only tentative because of the low S/N of the detections). No significant variations of the polarization position angle $\theta$ across the broad lines were detected. We also do not see any evidence for the broad lines to have polarization fractions any different from the surrounding continuum. Because of the low S/N of the polarized spectra (due to the faintness of the objects), the presence of a hidden broad-line region cannot be ruled out for any of the sources in which the broad lines have not been detected in polarized light. For these objects, our typical sensitivity to broad lines in the polarized light is about 100\AA\ in equivalent width for a line with a FWHM of 4000 km sec$^{-1}$. This limit is significantly above the equivalent width of the H$\beta$ line in the composite quasar spectrum (46\AA; \citealt{vand01}), so undetection of the broad lines in the polarized spectra is consistent with expected line intensities.

Narrow lines are detected in the polarized spectra of SDSS~J0842+3625 and SDSS~J1039+6430 (Figures \ref{pic1}-\ref{pic2}) and possibly of three more objects; all detected lines are listed in Table 1. Again, given the low S/N of our polarized spectra the presence of polarized narrow emission lines is not ruled out for any of the sources (the typical limiting equivalent widths are about 100\AA, similar to the sensitivity to the broad lines). But it can be stated with certainty that in all cases the fractional polarization in the strong narrow emission lines (specifically, [OIII]4959, 5007) is significantly less than that of the surrounding continuum. 
 
\subsection{Individual objects}
\label{sec_ind_obj}

\subsubsection{SDSS~J0842+3625}
\label{sec_ind_0842}

This object (Figure \ref{pic1}) has the highest [OIII]5007 luminosity in our sample \\($\log (L$([OIII]5007)/$L_{\odot})=10.1$; Paper I). It is the second most [OIII]5007-luminous type II quasar known at low redshifts ($z\la 1$), after IRAS09104 \citep{klei88, hine93, fran00, iwas01}. The polarization fraction is roughly wavelength-independent (excluding the narrow emission lines with very low fractional polarization) at an extraordinarily high level of 16\% over the entire range observed by SPOL. The polarization of SDSS~J0842+3625 is comparable to the highest polarization detected in any type II AGN (20\%; \citealt{anto85, hine93, tran95b, schm02}), and there are no discernible stellar features in its total optical spectrum. Therefore, dilution by unpolarized light (host galaxy or direct light from the AGN) is probably insignificant in this object. The broad H$\gamma$ emission line is tentatively detected in polarized light, as are the narrow H$\beta$ and the [OIII]4959, 5007 emission lines. Broad H$\beta$ is not detected because the A band of terrestrial O$_2$ absorption (7600\AA) makes it very difficult to detect spectral features at this wavelength. The power-law index of the polarized spectrum $\alpha_{\nu}=0.0\pm 0.2$ is similar to those of normal quasars (Section \ref{sec_pol_cont}). The polarization angle is constant over the entire range observed by SPOL except the region around [OIII]4959, 5007, where it rotates by about 20 degrees. The shape of the total optical continuum (Figure \ref{pic1}d) shows a break at rest-frame 3700\AA, both in the MMT spectrum and in the original SDSS spectrum of much lower S/N; we interpret this feature as the Balmer continuum in emission, presumably associated with the narrow-line region \citep{will85}.

\subsubsection{SDSS~J1039+6430}
\label{sec_ind_1039}

In this source, broad H$\beta$ and H$\gamma$ are both detected in the polarized spectrum (Figure \ref{pic2}), with the ratio of line fluxes H$\beta$/H$\gamma=1.4-1.8$. In addition, in the polarized spectrum we tentatively find excess emission over the best-fit power law between rest-frame 4450\AA\ and 4750\AA, which we interpret as the FeII complex \citep{vand01}. Over the observed spectral range, the polarization fraction declines by about a factor of two toward long wavelengths, but the host galaxy fraction at the longest wavelength is about half of the observed continuum light, so when corrected for dilution by stellar light, the polarization fraction is consistent with being wavelength-independent. The polarized spectrum power index is similar to those of normal quasars ($\alpha_{\nu}=0.0$, Section \ref{sec_pol_cont}). In Section \ref{sec_discussion} we discuss in detail the HST images of this object in connection with spectropolarimetric observations.

\subsubsection{SDSS~J1715+2807}
\label{sec_ind_1715}

Our spectropolarimetry for SDSS~J1715+2807 (Figure \ref{pic_broad}, right) is based on a total exposure time of 4.3 hours, a 1.8 hour increase over the exposure used in \citet{smit03}. This object is 2MASS J171559.7$+$280717 \citep{cutr02} selected based on its red infrared color ($J-K_S>2$); it was independently recovered from the SDSS data as a type II quasar candidate. \citet{smit03} suggested the possible presence of a very broad emission feature in the polarized spectrum at the position of H$\beta$, and with a longer exposure we confirm this suggestion. This feature may be interpreted either as a low S/N blend of the broad H$\beta$ and the [OIII]4959, 5007 emission lines, or a very broad ($FWHM=11400$ km sec$^{-1}$) redshifted (by about 4000 km sec$^{-1}$) H$\beta$ line. If the latter interpretation is correct, then the red shape of the polarized spectrum and the extreme properties of the broad emission line are remarkably similar to the properties of the unusual red AGN 2MASS J130005.3$+$163214 \citep{schm02}. We also confirm the presence of weak underlying broad components in H$\beta$ and H$\gamma$ in the total optical spectrum of SDSS~J1715+2807, which led \citet{smit03} to classify this object as a type 1.8 AGN rather than a `pure' type II AGN. 

The interpretation of this object, as discussed in Section \ref{sec_broad_band}, is that we are observing a heavily reddened AGN continuum: the continuum slope in the optical redward of 6000\AA\ and in the near-IR\footnote{Near-infrared photometry is from the 2MASS catalog available at the Infrared Science Archive, http://irsa.ipac.caltech.edu/.} can be roughly reproduced by applying rest-frame extinction with $A_V\simeq 3-4$ to a typical unobscured quasar continuum with $\alpha_{\nu}=-0.44$ \citep{vand01}, whereas in `true' type II quasars the obscuring material is optically thick even in the 2MASS bands. Similarly, using the optical continuum flux from \citet{schm02} and the 2MASS photometry for 2MASS~J130005 we find $A_V\simeq 5-5.5$. To estimate extinction values we used a Milky Way-like extinction curve as given by \citet{card89} and \citet{odon94}, and the uncertainty in the extinction value reflects the uncertainty in the selective absorption parameter ($2.5<R_V<5.0$; the extinction curves are not very sensitive to $R_V$ for $\lambda>5000$\AA), but we have not taken into account the possible deviations of the individual intrinsic spectra from the median. Using a high redshift quasar sample, \citet{hopk04} showed that reddening in quasars is better described by SMC-like extinction than by Milky Way-like extinction with $R_V=3.1$, but the two curves are very similar for $\lambda>2500$\AA, and all the differences arise in the rest-frame far-UV \citep{pei92}. For our extinction estimates we have considered only the optical rest-frame $\lambda>6000$\AA\ region and the near-IR bands, and therefore our conclusions are very insensitive to the assumptions about the extinction law. 

\subsubsection{SDSS~J1543+4935}
\label{sec_ind_1543}

This object (Figure \ref{pic_broad}, left) also shows evidence for a broad component in H$\beta$ in the total optical spectrum, and possibly a red excess similar to that seen in SDSS~J1715+2807. Given the low S/N in the red part of the spectrum, it is not clear whether the red excess in SDSS~J1543+4935 is due to the stellar light from the host galaxy or if it is the heavily reddened AGN continuum. This object is not detected in any of the 2MASS bands\footnote{We used the Quicklook Image service at the Infrared Science Archive, http://irsa.ipac.caltech.edu/, to look for possible low-significance matches not included in the 2MASS catalog.}, and since the optical continuum fluxes of SDSS~J1715+2807 and SDSS~J1543+4935 are very similar, as are their [OIII]5007 luminosities and redshifts, the latter object must be more strongly obscured than the former. 

\subsubsection{SDSS~J1501+5455}
\label{sec_ind_1501}

This is the only object in the sample in which the luminosity of the host galaxy is significantly above $L^*$, as estimated based on the contribution of the host to the total optical spectrum. The SDSS image shows a resolved elliptical galaxy with a maximum extent in the $i$-band of about 12\arcsec, corresponding to a physical size of 58 kpc; thus using the spectroscopic flux for calculating the host luminosity is inappropriate since the spectrum is obtained in a 3\arcsec\ aperture (SDSS fiber diameter). Using K-corrected Petrosian magnitudes from the SDSS photometric pipeline, we find $M_r=-24.1$, which can be compared to the value of $M^*=-20.8$ in the $r$-band obtained by \citet{blan01} for the luminosity function of SDSS galaxies. Therefore the luminosity of the host galaxy of SDSS~J1501+5455 is $\sim 21 L^*$, and the host is almost certainly a cD galaxy. This object is the only radio-loud AGN in the sample presented in this paper; it shows an unusual radio morphology and has a high soft X-ray luminosity (Paper II, Appendix A4). The measured polarization is rather low (0.3-1.8\%), but the continuum is completely dominated by the bright host galaxy ($h_y>0.8$, Table 1), so the host-corrected polarization is much higher.

\section{Discussion}
\label{sec_discussion}

\subsection{Polarization mechanism}

Only a limited number of mechanisms can produce light with a fractional linear polarization of up to 20\%. In this section we briefly review these mechanisms and discuss their pros and cons as explanations for the polarization of SDSS type II quasars in this paper.

{\bf Synchrotron emission} has been long ruled out as the primary mechanism of optical emission in most AGNs (except blazars, \citealt{urry95}). This view is supported by the spectropolarimetric observations: (i) this mechanism cannot explain the broad lines or narrow lines in the polarized spectra; (ii) the fractional polarization would be the same for type I and type II AGNs if the continuum light was moderately extincted in type II AGNs, and (iii) type II AGNs would not be polarized at all if the continuum emission was heavily extincted. 

{\bf Dichroic extinction} by intervening dust produces linearly polarized light because non-spherical dust grains aligned in a large-scale magnetic field extinct one incident polarization more strongly than the other \citep{drai03a}. The maximum degree of polarization is related to the amount of dust and therefore to the reddening: $q\la 9\% \times E(B-V)$ \citep{serk75}. Dichroic extinction produces a wavelength-independent polarization angle which reflects the direction of grain alignment. Since this mechanism requires large extinction values to produce a polarization of several percent, it is usually ruled out for highly polarized sources with a blue polarized continuum (e.g., \citealt{jann94} and most of the objects in our sample, e.g. Figures \ref{pic1}-\ref{pic2}). 

However, dichroic extinction can explain polarization in the small class of objects with red, not blue, polarized spectra -- SDSS~J1715+2807 (this paper and \citealt{smit03}) and 2MASS~J130005 \citep{schm02, smit03}. The much smaller fractional polarization in the narrow emission lines in both these objects is consistent with the narrow lines originating above the bulk of the obscuring material. To produce the observed polarization level in SDSS~J1715+2807 by dichroic extinction would require $A_V$ of 3 or 4 magnitudes, similar to what is needed to produce the observed red continuum by reddening a typical quasar spectrum (Section \ref{sec_ind_1715}). Similarly, the rest-frame extinction required to produce the observed optical and near-IR slope in 2MASS~J130005 (from \citealt{schm02} and 2MASS photometry), if applied to a typical unobscured quasar continuum, is about $A_V=5$ magnitudes, enough to produce a dichroic polarization of up to 15\%. 

We emphasize that these estimates are based on observations of interstellar polarization in the Galaxy, which might not be directly applicable to extragalactic objects or to high extinction values. In particular, we extrapolated the Serkowski law to much higher reddening values than those for which the dichroic extinction has been observed. Therefore, our conclusions about the possibility of strong dichroic polarization in the two red AGNs are only suggestive. An additional difficulty is posed by the very high fractional polarization of the H$\alpha$ line in 2MASS~J130005, which is probably too high to be accommodated by dichroic extinction \citep{schm02}. 

{\bf Scattering} by off-nucleus electrons or dust is the widely assumed mechanism for producing polarization in most polarized AGNs. This mechanism is strongly supported by the direct detection of the scattering material in nearby objects through imaging polarimetry (e.g., \citealt{anto85, kish99}). 

\begin{figure}
\epsscale{0.5}
\plotone{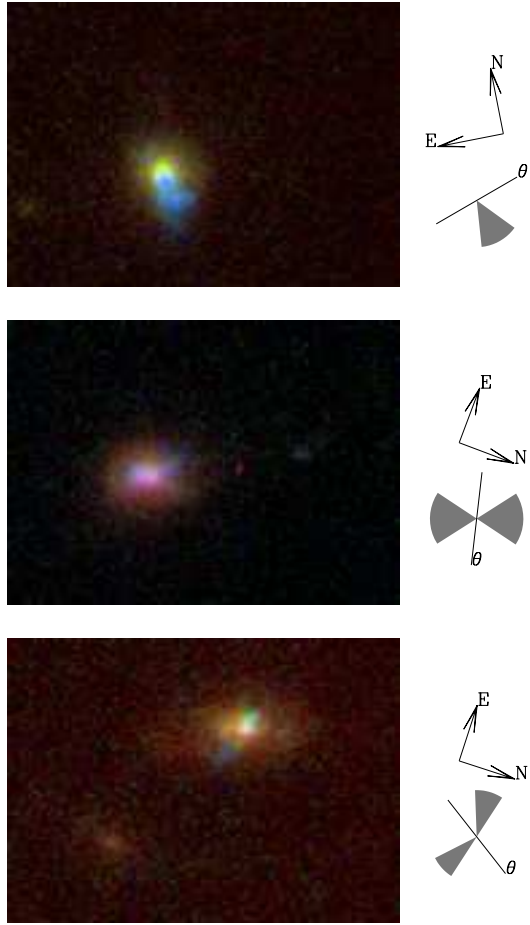}
\figcaption{\small Three-band HST color-composite images of SDSS~J1039+6430 (top), SDSS~J1323$-$0159 (middle) and SDSS~J1413$-$0142 (bottom). The orientations on the sky are indicated with arrows on the right of the images. All images are 146$\times$106 pixels or 7.3\arcsec$\times$5.3\arcsec. This translates into physical sizes of 39$\times$29 kpc, 36$\times$26 kpc and 38$\times$28 kpc, correspondingly. For SDSS~J1039+6430 (top), the irregularly shaped blue spot is identified as a one-sided scattering region. In SDSS~J1323$-$0159 (middle) the scattering region has a fairly symmetric bi-conical shape. In SDSS~J1413$-$0142 the scattering structure is somewhat asymmetric and is seen on both sides of the nucleus. Schematic representations of the scattering regions are given on the right of each figure in grey using position angles listed in Table 2 (rotated into the coordinate systems of the HST images). The orientations of the polarization planes are indicated by solid lines labeled $\theta$. The scattering regions are orthogonal to the polarization position angles in all three cases within the errors. \label{pic_hst}}
\end{figure}

While our objects are much more distant and cannot be mapped to a similar level of detail, for the three objects in our sample with HST observations (Table 1) we still can identify the scattering regions in our three-band ACS images shown in Figure \ref{pic_hst}, even though our program did not include imaging polarimetry. The blue regions in SDSS~J1323$-$0159 and SDSS~J1413$-$0142 are very similar to scattering cones in nearby AGNs \citep{pogg93}, and we interpret them as scattering regions by analogy. This morphological similarity alone strongly suggests scattering as the main polarization mechanism in these objects. In SDSS~J1039+6430, the blue region has a more unusual shape, but we still interpret most of this light as scattered emission rather than light from young stars, given that the polarization fraction in the blue is 20\% (Figure \ref{pic2}), similar to the highest polarization ever measured for type II AGNs. The position angles of the scattering cones in the three objects with HST observations are given in Table 2. We emphasize again that the bands for HST imaging were carefully selected to avoid bright emission lines, and thus the HST images show continuum emission almost exclusively. Therefore, in these images we are seeing scattering cones rather than ionization cones. 

For two of these objects (SDSS~J1039+6430 and SDSS~J1323$-$0159), we find that the orientation of the scatterers is perpendicular within 5$^{\circ}$ to the position angle of the polarization, in excellent agreement with what is expected from scattering. The scattering regions are somewhat asymmetric in SDSS~J1413$-$0142 relative to the nucleus, but they are also perpendicular to the polarization plane, within the measurement error. Thus, combined MMT and HST observations strongly support scattering as the dominant polarization mechanism in these three objects.

SDSS~J1715+2807 was also imaged by HST \citep{marb03}, but only in the $I$ band, which is dominated by the light of the host galaxy, with a possible contribution from the direct reddened light of the AGN. Because of the lack of rest-frame UV observations, the extent and orientation of the scattering region are unknown in this object. 

The small fractional polarization of the narrow emission lines and the variation of the position angle across them found in some objects indicate that the size of the scattering region is comparable to the size of the narrow-line region. While we do not have any direct observations of the size of the narrow-line regions in the objects from our sample, the narrow-line regions in radio-loud type II quasars with similar [OIII]5007 luminosities extend out to several kpc and several tens of kpc in some cases \citep{mcca93}. The extents of the narrow-line regions are therefore similar to the sizes of the scattering regions inferred from the HST images (several kpc, up to 10 kpc in the case of SDSS~J1039+6430, Figure \ref{pic_hst}), in support of the scattering picture.

\subsection{Scattering: dust vs electrons}
\label{sec_dust_el}

Despite extensive observations, distinguishing between dust scattering and electron scattering has proved to be very difficult, and an extensive literature exists on comparing the two mechanisms (e.g., \citealt{kish01} and references therein). The two spectroscopic diagnostics are the scattering efficiency (the scattered flux density divided by the flux density we would have detected in a direct view to the central source) and the polarization fraction of the scattered light (the polarized flux density divided by the scattered flux density). Both of these values are dependent on the host galaxy correction that is applied to separate the scattered component from the total optical spectrum (the contribution of the direct light from the AGN is unimportant in most type II quasars). In addition, the derived scattering efficiency depends on the assumed incident spectrum, which in the case of type II AGNs is obscured and unknown to the observer. 

Electron scattering produces wavelength-independent scattering efficiency and polarization fraction. In type II AGNs, where the contribution of the direct light from the nucleus is negligible, electron scattering would produce (i) a polarized flux density that has exactly the same spectral shape as that seen by the scatterer and should be very similar to a typical unobscured AGN; and (ii) an observed polarized fraction that is wavelength-independent when corrected for unpolarized stellar light. These properties are found in many type II AGNs, including some objects in our sample (Table 1 and Figures \ref{pic1}-\ref{pic2}). On the basis of spectral properties, electron scattering has been suggested as the primary polarization mechanism in NGC1068 \citep{anto85, mill91}. 

Dust scattering generally produces a wavelength-dependent polarization fraction and scattering efficiency \citep{zubk00, drai03b}, and the strongest wavelength dependence of these parameters is in the far-UV. This spectral region can be probed in high-redshift sources \citep{hine01, kish01} when it is redshifted into the near-UV and optical. In these cases spectropolarimetry alone can provide strong constraints on the nature of the scattering agent and scattering geometry. In our work we probe rest wavelengths $\lambda>2500$\AA. At these wavelengths optically thin dust scattering is not always strongly wavelength-dependent, but typically it produces (i) a scattering efficiency rising to the blue and (ii) a polarization fraction of the scattered light rising to the red \citep{zubk00, witt00, drai03b}.  With the exception of the unusual object SDSS~J1715+2807 (Section \ref{sec_ind_1715}), red polarization fractions are not observed in our sample, and the total optical spectra of our objects are no bluer than typical quasar continua. 

The wavelength dependence of dust scattering can be strongly affected by the dust composition \citep{drai03b}, geometry of the scattering region \citep{hine01} or dust reddening at any stage of the light propagation \citep{schm02}. In addition, multiple-scattering models can accommodate a large variety of wavelength dependences \citep{manz96, zubk00, hine01, kish01, kish02}. For example, multiple scattering and absorption in opaque clouds can produce an apparently gray scattering \citep{kish01}. The dilution-corrected polarization values of 10\%-20\% typical for type II AGNs agree better with dust scattering than with electron scattering, which would produce typically higher polarization levels (e.g., \citealt{cima94}). Observationally, the wavelength dependence of the scattering efficiency is difficult to determine because of the uncertainty in the host galaxy contribution. In other words, in practice it is often impossible to distinguish between dust scattering and electron scattering based on the rest-frame optical spectral properties alone, and other considerations must be applied.

One such consideration is that electron scattering is often difficult to reconcile with the size of the scattering regions if they are significantly larger than a few hundred parsecs \citep{dise94, cima96, dey96, hine01, kish01}. We illustrate this argument for SDSS~J1039+6430, which has the largest scattering region of the three objects in our sample with HST imaging observations. 

The scattering efficiency $\varepsilon$ is the ratio of the detected scattered flux density to the flux density we would detect if we had a direct view to the source; it can be calculated if the size of the scattering region and the density of the scattering particles are known:
\begin{equation}
\varepsilon(\upsilon)=\frac{d\sigma(\upsilon)}{d\Omega}\langle n\rangle V \left\langle\frac{1}{d^2}\right\rangle, \label{eq_ne}
\end{equation}
where $d\sigma(\upsilon)/d\Omega$ is the differential cross-section as a function of the scattering angle $\upsilon$ (for electrons scattering at 90$^{\circ}$ it is $7.9\times 10^{-26}$ cm$^2$ sr$^{-1}$), $\langle n\rangle$ is the volume-averaged density and $\langle 1/d^2\rangle$ is the inverse square distance of each scattering particle from the illuminating source averaged over all particles (approximately equal to $3/d_{max}^2$ for conical geometry and density independent of $d$). The volume of the scattering region is $V=d_{max}^3\Delta\Omega/3$, where $\Delta\Omega$ is the solid angle subtended by the scatterer as seen from the central source, and it can be estimated from the HST images and Table 2; for example, for SDSS~J1039+6430 $d_{max}\simeq 10$ kpc and $\Delta\Omega=0.5$ sr (calculated assuming that the axis of the cone lies in the plane of the sky). The scattering efficiency is very poorly constrained since we do not know the intrinsic luminosity of the source, but if we compare obscured and unobscured quasars with similar [OIII]5007 luminosity (similar to our approach in Paper I) we find that we observe about $\varepsilon=1\%$ of typical unobscured quasar flux as scattered light. For the electron scattering to provide such high efficiency, equation (\ref{eq_ne}) then implies an average electron density of 8 cm$^{-3}$ for the parameters of SDSS~J1039+6430. The resulting emission measure is 
\begin{equation}
EM\equiv\langle n^2 \rangle V=3.1\times 10^{68}\mbox{ cm}^{-3}\cdot\frac{\kappa^2}{\sin^2\upsilon}\left(\frac{d_{max}}{10\mbox{ kpc}}\right)\left(\frac{\Delta\Omega}{0.5\mbox{ sr}}\right)^{-1}\left(\frac{\varepsilon}{1\%}\right)^2, \label{eq_EM}
\end{equation}
where $\kappa\equiv\sqrt{\langle n^2\rangle}/n$ is the `clumping factor' which by definition is $\ge 1$ and can be rather greater. Furthermore, if the axis of the scattering cone does not lie in the plane of the sky, then we underestimated the extent of the scattering region, and $d_{max}/\Delta\Omega\propto \cos^{-3} i$, where $i$ is the inclination angle of the axis of the cone ($i=0^{\circ}$ for the plane of the sky). 

Given such a large value of $EM$, scattering off electrons with characteristic temperatures of about $10^4$ K is ruled out for SDSS~J1039+6430 because the resulting recombination line luminosity would be prohibitively large:
\begin{equation}
L({\rm H}\beta)= EM \times \frac{\hbar c}{2 \lambda({\rm H}\beta)}\alpha({\rm H}\beta) \frac{2X}{1+X}=6.1\times 10^8 L_{\odot} \cdot \frac{\kappa^2}{\sin^2\upsilon\cos^3 i}, \label{eq_line}
\end{equation} 
where $\alpha({\rm H}\beta)=3\times 10^{-14}$ cm$^3$ s$^{-1}$ is the recombination coefficient at temperature $10^4$ K and $X=0.7$ is the hydrogen fraction by mass. The observed H$\beta$ luminosity in this object is about $1.3\times 10^8 L_{\odot}$, and therefore estimate (\ref{eq_line}) is at least a few times higher than observed (the last term in eq. \ref{eq_line} can be much greater than unity). 

Scattering off electrons with characteristic temperatures of about $10^6-10^7$ K is ruled out by X-ray observations. The bremsstrahlung luminosity of a plasma with temperature in this range and $X=0.7$ is \citep{rybi79}
\begin{equation}
L_{br}=2\times 10^{-24} \, T_6^{1/2} \mbox{ erg cm$^3$ sec$^{-1}$} \times EM =6.2\times 10^{44} \, T_6^{1/2} \mbox{ erg sec$^{-1}$} \cdot \frac{\kappa^2}{\sin^2\upsilon\cos^3 i}, 
\end{equation} 
where $T_6=T/10^6$ K. Much of the bremsstrahlung emission would be emitted in the range $0.2-2$ keV for temperatures $2\times 10^6-10^7$ K. Such luminosities in the soft X-ray band are ruled out since the source is undetected in a ROSAT 700 sec exposure\footnote{We used the archive services and the tool WebPimms from the High-Energy Science Archive, http://heasarc.gsfc.nasa.gov/}, placing an upper limit on its $0.2-2.0$ keV luminosity of $L_X<6 \times 10^{43}$ erg sec$^{-1}$. There is no significant reddening of the scattered light, so there cannot be any significant extinction of the putative bremsstrahlung emission.

Scattering off electrons with characteristic temperatures around $10^5-10^6$ K is not constrained by similar arguments. However, an independent constraint comes from observation of narrow emission lines in the polarized spectrum of SDSS~J1039+6430 (Figure \ref{pic2}) which would be washed out if they were produced by scattering off hot electrons \citep{fabi89, cima96}. Specifically, the observed widths of narrow lines are due to the resolution of the spectrograph, the intrinsic dispersion and thermal broadening of the scattering particles: $\sigma_{obs}^2=\sigma_{sp}^2+\sigma_{int}^2+\sigma_{th}^2$, where thermal broadening is related to the electron temperature as $\sigma_{th}^2\sim 2 k T_e/m_e$. The S/N of the polarized spectrum does not allow detailed line fitting of the [OIII]4959,5007 emission lines, but the observed $FWHM$([OIII]) is constrained to be $<3000$ km sec$^{-1}$, dominated by $\sigma_{sp}$, so that $\sigma_{th}<$ 600 km sec$^{-1}$, and scattering off electrons warmer than $10^4$ K is ruled out in this object.

If, on the contrary, we assume scattering by dust, the cross-section of scattering is about $10^{-24}-10^{-23}$ cm$^2$ sr$^{-1}$ per H nucleon for different dust compositions \citep{drai03b}, and equation (\ref{eq_ne}) implies a hydrogen density of about $n_H\simeq 0.6-6$ cm$^{-3}$. Extinction by that same dust over a distance $d$ is important but not prohibitive, with 
\begin{equation}
A_V=0.17\times \left(\frac{\sigma_{ext}(V)}{5\times 10^{-23}\mbox{ cm$^2$ H$^{-1}$}}\right) \left(\frac{n_H}{1 \mbox{ cm$^{-3}$}}\right) \left(\frac{d}{1\mbox{ kpc}}\right),
\end{equation}
where $\sigma_{ext}(V)$ is the extinction cross-section in the $V$-band (following \citealt{hopk04}, we used SMC-like extinction values from \citealt{drai03b}). 

Finally, in SDSS~J1039+6430 the ratio H$\beta$/H$\gamma=1.4-1.8$ (calculated based on best-fit Gaussians) is less than that of the composite spectrum of SDSS quasars (H$\beta$/H$\gamma=3.3$, \citealt{vand01}) or the Case B recombination value (H$\beta$/H$\gamma=2.0-2.2$, depending on the temperature, \citealt{oste89}). This effect cannot be attributed to reddening, which would increase the H$\beta$/H$\gamma$ ratio. This measurement suggests that the polarized spectrum is bluer than the incident spectrum, inconsistent with electron scattering. 

Similar arguments can be applied to the other two objects in our program with HST observations, SDSS~J1323$-$0159 and SDSS~J1413$-$0142. The extents of the scattering regions are not quite as impressive in these two objects as in SDSS~J1039+6430, but they are still several kpc in each dimension, which would result in large values of $EM$. Scattering off electrons at temperatures $10^4$ K and $>10^6$ K would be ruled out because of the prohibitively large line emission or X-ray emission (neither of these objects is detected by ROSAT). However, narrow emission lines are not observed in the polarized spectra of either of these two objects, so the broadening argument to rule out electron scattering at intermediate temperatures ($10^5-10^6$ K) does not apply. 

\subsection{Optical selection efficiency of type II quasars}
\label{sec_ind_efficiency}

The S/N ratios of the total optical spectra obtained with MMT are much higher than those of the original SDSS spectra, which enables us to study weak emission features. For example, the MMT spectra of four objects show emission lines on both sides of the MgII 2800 doublet (as illustrated in Figure \ref{pic1}d for SDSS~J0842+3625) which we identify as HeII 2733.3 and CII 2836.7, 2837.6\footnote{High-precision emission line wavelengths in the air were taken from the Atomic Line List hosted by the Department of Physics and Astronomy at the University of Kentucky, http://www.pa.uky.edu/$\tilde{\,\,\,}$peter/atomic.} \citep{kinn93}. The blend of these emission lines with MgII 2800 could be mistaken for a broad component in the MgII 2800 emission line in low S/N spectra and/or at low spectral resolution. The ratios HeII 2733.3 / HeII 4686 and HeII 2733.3 / HeII 3202 measured for SDSS~J0842+3625 are roughly 0.2 and 0.5, consistent with Case B recombination line ratios \citep{oste89}. Similarly, we interpret what looks like a broad bump underlying the HeII 4686 emission line (Figures \ref{pic1}d and \ref{pic2}d) as a blend of OIV$+$SV 4654, HeII 4686 and a doublet [ArIV] 4711.3, 4740.2. This blend is observed in six of our objects. In the remaining spectra (e.g., SDSS~J1715+2807, Figure \ref{pic_broad}) only the HeII 4686 line is observed, without the accompanying high-ionization lines. 

The mere presence of a broad component in an emission line in the total spectrum does not by itself indicate that an object cannot be classified as a type II AGN -- the entire broad line could be scattered. For example, to produce the broad component in SDSS~J1543+4935 entirely by scattering, an efficiency of 1.3\% would be required (assuming that the unobscured broad H$\beta$/[OIII]5007 ratio is similar to that of the quasar composite spectrum, \citealt{vand01}). No direct measurement of scattering efficiency exists, but values of about 1\% are suggested for our objects by the luminosity argument, as described in Section \ref{sec_dust_el}. The absence of a near-IR detection of this object (Section \ref{sec_ind_1543}) suggests that the broad line is scattered light rather than highly reddened direct light from the broad-line region. 

We mentioned in Paper I that the optical identifications of our sources as type II AGNs are uncertain due to the quality of the SDSS spectra, as we could have missed underlying broad components in emission lines if their amplitudes were comparable with the noise. The much deeper spectroscopy for 12 of the sources presented in this paper shows evidence for weak broad components in H$\beta$ in two of them (SDSS~J1543+4935 and SDSS~J1715+2807; \citealt{smit03}). Of these two objects one (SDSS~J1715+2807) is a heavily reddened type 1.8 AGN. The broad component in SDSS~J1543+4935 is likely to be entirely due to the scattered light, and since the direct light from the broad-line region is obscured, it can be classified as a type II AGN. We do not see any evidence for broad components in any other emission lines (after we identified possible blends as described above). This observation indicates that the optical classifications of high-luminosity type II AGNs in Paper I are valid for roughly 90$-$95\% of the sample. 

Objects with high scattering efficiencies (a few per cent or larger) can have broad components that are strong enough to be identifiable in the original lower S/N SDSS spectra. Such objects would have been rejected by our selection procedure (described in Paper I). We created simulated spectra of type II AGNs with luminosities and redshifts typical of our sample, applied a range of scattering efficiencies and typical SDSS spectroscopic noise and found that if the scattering efficiency was less than 3\%, broad components were too weak to be identified in the SDSS spectra. 

Finally, since 10 of the objects observed with MMT show no underlying broad components in the total spectra, the scattering efficiencies in these objects are $\varepsilon\la 1\%$. This implies that we can place lower limits on the intrinsic luminosities of the objects in our sample by measuring the scattered luminosities. For example, the SDSS spectrum of SDSS~J0842+3625 at 4000$-$6000\AA\ (rest-frame 2560$-$3840\AA) implies a total flux in this range of $8\times 10^{-14}$ erg sec$^{-1}$ cm$^{-2}$, and therefore the scattered luminosity in this spectral region is $1\times 10^{44}$ erg sec$^{-1}$. The condition $\varepsilon\la 1\%$ implies that the intrinsic luminosity of the object in SDSS~J0842+3625 is $\ga1\times 10^{46}$ erg sec$^{-1}$.

\section{Conclusions}
\label{sec_conc}

Spectropolarimetry of AGNs offers the possibility of observing their central regions along multiple lines of sight. Some of the light from the AGN is scattered off the surrounding material and is therefore polarized. In type II AGNs the direct light from the central parts of the AGN is so heavily extincted that spectropolarimetry provides the only way to explore the central engine at optical wavelengths. 

Out of the 12 type II quasar candidates in this study, nine show high polarization ($P>3\%$). If a crude host galaxy correction is applied then all objects are highly polarized by this criterion. The optical continua of most objects are dominated by the highly polarized scattered light and the unpolarized host galaxy component; the direct light from the AGNs is a negligible contribution except for one objects which is also a red 2MASS AGN and is atypical of our sample. This conclusion is reinforced by the absence of near-IR excesses for most of our objects (Paper II). 

Broad emission lines were detected in five objects in the polarized spectra. This is direct evidence that type II quasar candidates from the SDSS harbor central regions similar to those of type I AGNs, with the defining characteristics of a blue continuum and broad emission lines. The presence of the broad-line region is not ruled out for any of our objects due to the low S/N of the spectropolarimetry of these faint objects, with a typical sensitivity to broad lines of about 100\AA\ in equivalent width. 

Three objects with spectropolarimetric observations have also been imaged in three bands as part of our HST program (SDSS~J1039+6430, SDSS~J1323$-$0159 and SDSS~J1413$-$0142, Figure \ref{pic_hst} and Paper IV). For these objects, the combination of spectropolarimetry and HST imaging strongly favors scattering as the mechanism producing polarization, since in all three cases the plane of polarization is orthogonal to the orientation of the presumed scattering regions. The wavelength dependence of the scattering efficiency can in some cases be used to discriminate between dust and electron scattering, but it is difficult to determine observationally because of the highly uncertain host galaxy fraction. We use other arguments to show that the object SDSS~J1039+6430 is a strong candidate for dust scattering. In addition to having a very large scattering region ($\sim$10 kpc) it shows narrow emission lines in the polarized spectrum, which rules out scattering by warm and hot electrons. In this object there is also evidence that the polarized spectrum is bluer than the intrinsic spectrum, inconsistent with wavelength-independent electron scattering. In the remaining two objects SDSS~J1323$-$0159 and SDSS~J1413$-$0142 the scattering regions as seen in the blue band of the HST images have extents over 1 kpc, but scattering by warm electrons ($\sim 10^5$ K) is not excluded as the main polarization mechanism.  

All of these observations can be well described by a simple unification model with obscuration in a plane and scattering by material outside this plane. The size of the scattering region is comparable to the size of the narrow-line region, explaining the low polarization of the narrow-line emission if the narrow-line gas is intermixed with the scattering material. Typical extinction values in the obscuring plane should be well in excess of $A_V=7$. We suggest that the typical scattering efficiency is around 1\%.  

One out of 12 objects in our program (SDSS~J1715+2807) shows a non-stellar red continuum, polarization fraction rising to the red and a red polarized spectrum. This object is also a red 2MASS AGN, and it is better understood as a heavily reddened quasar rather than a completely obscured type II AGN \citep{smit03}. We suggest that polarization by dichroic extinction in the obscuring torus with $A_V\simeq 4$ cannot be excluded as the primary polarization mechanism in this object.

Finally, SDSS~J0842+3625, the most luminous type II quasar in our sample, shows an extraordinary wavelength-independent polarization level of 16\%. A broad H$\gamma$ line is detected in the polarized spectrum, as well as narrow H$\beta$ and [OIII]4959, 5007. We estimate that the intrinsic luminosity of this object in the near-UV (rest-frame 2560$-$3840\AA) is at least 10$^{46}$ erg sec$^{-1}$. 

\section*{Acknowledgments}

Funding for the creation and distribution of the SDSS Archive has been provided by the Alfred P. Sloan Foundation, the Participating Institutions, the National Aeronautics and Space Administration, the National Science Foundation, the U.S. Department of Energy, the Japanese Monbukagakusho, and the Max Planck Society. The SDSS Web site is http://www.sdss.org/. 

The SDSS is managed by the Astrophysical Research Consortium (ARC) for the Participating Institutions.  The Participating Institutions are The University of Chicago, Fermilab, the Institute for Advanced Study, the Japan Participation Group, The Johns Hopkins University, the Korean Scientist Group, Los Alamos National Laboratory, the Max-Planck-Institute for Astronomy (MPIA), the Max-Planck-Institute for Astrophysics (MPA), New Mexico State University, University of Pittsburgh, Princeton University, the United States Naval Observatory, and the University of Washington.

Based on observations made with the NASA/ESA Hubble Space Telescope as part of the program 9905. Support for this work was provided in part by NASA through grant HST-GO-09905.01 from the Space Telescope Science Institute (STScI). STScI is operated by the Association of Universities for Research in Astronomy, Inc., under NASA contract NAS 5-26555. 

NLZ and MAS acknowledge the support of NSF grant AST-0307409. NLZ also acknowledges the support of the Charlotte Elizabeth Procter Fellowship. PSS acknowledges support from NASA grant 1256424.

The authors would like to thank Bruce Draine for useful discussions and the referee for helpful suggestions.

\clearpage

\clearpage

\clearpage

\clearpage

\clearpage

\clearpage
\begin{deluxetable}{cccccccccccccc}
\pagestyle{empty}
\rotate
%\tabletypesize{\scriptsize}
\tabletypesize{\tiny}
\tablewidth{0pt}
\setlength{\tabcolsep}{0.03in}
\tablecaption{Spectropolarimetry of Type II Quasars}
%\tablewidth{475.00000pt}
%\tablenum{1}
\tablehead{J2000 & redshift, & & radio  & exp. time, & mean & blue & green & yellow & mean $\theta$, & $FWHM$, & narrow & host & \\
coordinates & z & $L$([OIII]5007) & loud? & sec & polarization & polarization & polarization & polarization & deg. from north & km sec$^{-1}$ & lines & fraction & comments}
\startdata
081507.42$+$430427.2 & 0.510   & 9.57  & no & 3200  & 5.0$\pm$0.4 &  8.5$\pm$0.6 &  3.1$\pm$0.4 &  1.3$\pm$1.1 &  166$\pm$2 &  & & 0.6 & \\
084234.94$+$362503.1  & 0.561  & 10.10  & no  & 9600  & 15.4$\pm$0.7 & 16.5$\pm$0.2 & 16.2$\pm$0.3 & 14.9$\pm$1.8 & 11$\pm$1 &  4900 (H$\gamma$)& MgII, [OIII], H$\beta$ & 0 &  \\
103951.49$+$643004.2  & 0.402   & 9.41  & no  & 3200  & 10.9$\pm$0.2 & 16.6$\pm$0.3 & 10.2$\pm$0.2 &  8.0$\pm$0.3 & 109$\pm$1 & 3070 (H$\beta$) & [OIII] & 0.5 & HST \\
& & & & & & & & & & 4030 (H$\gamma$) & & & \\
132323.33$-$015941.9 & 0.350 & 9.19 & no & 3200 & 3.7$\pm$0.4 &  5.0$\pm$1.0 &  3.7$\pm$0.5 &  2.8$\pm$0.5 & 104$\pm$3 & & & 0.5 & HST \\
140740.06$+$021748.3 & 0.309 & 8.90 & no & 3200 & 1.9$\pm$0.4 &  2.6$\pm$0.2 &  0.8$\pm$0.4 &  1.0$\pm$0.4 & 142$\pm$6 & & & 0.8 & \\
141315.31$-$014221.0 & 0.380 & 9.25 & no & 3200 & 2.7$\pm$0.7 &  4.1$\pm$1.0 &  2.9$\pm$0.4 &  1.2$\pm$0.6 & 146$\pm$8 & & & 0.6 & HST \\
150117.96$+$545518.3 & 0.338 & 9.06 & yes  & 1920 & 0.4$\pm$0.1 & 1.8$\pm$0.3 &  0.6$\pm$0.1 &  0.3$\pm$0.1 & 121$\pm$10 & & & 0.9 & \\
150608.09$-$020744.2 & 0.439 & 9.25 & no & 3200 & 6.3$\pm$0.7 &  9.7$\pm$1.1 &  4.8$\pm$0.7 &  5.5$\pm$1.3 & 21$\pm$3 & & [OIII]?, H$\beta$? & 0.8? & \\
151711.47$+$033100.2 & 0.613 & 9.36 & no & 6440 & 7.0$\pm$0.6 & 10.5$\pm$0.8 &  6.7$\pm$1.2 &  n/a & 117$\pm$2 & & [OIII]? & 0.3(green) & \\
154340.02$+$493512.6 & 0.512 & 9.13 & no & 6440 & 5.8$\pm$0.2 & 7.7$\pm$0.3 &  6.1$\pm$0.2 &  3.2$\pm$0.7 & 127$\pm$1 & 4300 (H$\beta$) & & 0.4 & \\
164131.73$+$385840.9 & 0.596 & 9.92 & no & 3200 & 4.6$\pm$0.3 & 4.5$\pm$0.3 &  3.9$\pm$0.3 &  n/a & 13$\pm$2 & 3200 (H$\gamma$) & [OIII]?, H$\beta$? & 0.3(green) & \\
171559.79$+$280716.8 & 0.524 & 9.17 & no & 15520 & 4.6$\pm$0.2 & 3.5$\pm$0.3 &  6.0$\pm$0.2 &  5.6$\pm$0.4 & 6$\pm$1 & 11400? (H$\beta$) & & 0.2 & 2MASS
\enddata

\tablecomments{J2000 coordinates, redshifts and [OIII]5007 luminosities are from Paper I. [OIII]5007 luminosities are given as log($L$([OIII]5007)/$L_{\odot}$). Radio-loudness is determined as in Paper II, based on the $L$([OIII]5007)/$L(1.4\rm GHz)$ ratio. Exposure times refer to the MMT SPOL observations. All polarization values are in \%. `Mean polarization' is the average over the entire observed range (uncorrected for dilution). `Blue', `green' and `yellow' values are mean continuum polarization in the rest-frame 2820\AA$-$3710\AA, 4000$-$4840\AA\ and 5030$-$6530\AA\ bands, correspondingly (uncorrected for dilution; emission lines excluded). The quoted errors are 1$\sigma$ based on the observed pixel-by-pixel variations. $FWHM$ is the full width at half maximum of the broad emission lines in the polarized spectra based on the Gaussian fitting (tentative due to low S/N). The next column lists narrow lines detected in the polarized spectra (MgII is for MgII 2800 and [OIII] is for [OIII]4959,5007). The estimate of the host fraction is given for the `yellow' range (except SDSS~J1517+0331 and SDSS~J1641+3858). The most uncertain determination of the host galaxy fraction is for the spectrum of SDSS~J1506$-$0207, in which our estimated host fraction in the yellow band is uncertain by more than a factor of 2 (from $h_y=0.32$ to $h_y=0.75$). In the last column we indicate three objects for which HST imaging observations are available (Section \ref{sec_discussion} and Paper IV). SDSS~J1715+2807 is a red AGN 2MASS J171559.7$+$280717\citep{cutr02} that was independently recovered from the SDSS data as a type II quasar candidate; \citet{smit02p,smit03} presented first results of spectropolarimetry of this object and \citet{marb03} presented the HST $I$-band photometry of its host galaxy. The emission feature in the polarized spectrum is tentatively identified as H$\beta$ (see text for details). }
\end{deluxetable}

\clearpage
\begin{deluxetable}{cccc}
\tabletypesize{\small}
\pagestyle{empty}
\tablewidth{0pt}
\setlength{\tabcolsep}{0.1in}
\tablecaption{Position angles of scattering cones from the HST images}
\tablehead{J2000 & cone & cone & polarization \\
coordinates & 1 & 2 & position angle}
\startdata
103951.49$+$643004.2 & 175$-$223(N) & n/a & 109$\pm$1 \\
132323.33$-$015941.9 & 347$-$53(N) & 167$-$233(S) & 104$\pm$3 \\
141315.31$-$014221.0 & 75$-$110(E) & 225$-$255(W) & 146$\pm$8
\enddata
\tablecomments{This table includes data on three objects with HST images. For each scattering cone, the range of position angles it covers is given in degrees E of N. Cones are identified in the table by their position relative to the nucleus (N -- North, E -- East, S -- South, W -- West, see Figure \ref{pic_hst}). Polarization position angles are taken from Table 1.}
\end{deluxetable}

\end{document}